# Parametric Control of Nonlinear Longitudinal-Rotational Rod Oscillations and Phenomenon of Reverse Rotational Vibrations

Ivan V. Kazachkov[1,2]

[1]Dept of Information Technologies and Data Analysis,
Nizhyn Gogol State University, UKRAINE, http://www.ndu.edu.ua

[2]Dept of Energy Technology, Royal Institute of Technology, Stockholm, 10044, SWEDEN,
Ivan.Kazachkov@energy.kth.se, http://www.kth.se/itm/inst?l=en_UK

**Abstract:** The article is devoted to the investigation of the nonlinear effects in a system of the coupled longitudinal-torsional parametric vibrations of a rotating rod. Constructed and investigated mathematical model, based on which we calculated the resonance conditions of the nonlinear oscillations and found the ratio of the parameters that require changing of the sign of the coefficient in front of the term defining the possible reverse of torsional vibrations. The latter is a new phenomenon, a special mode, where the parametric action in the form of longitudinal vibrations at one end of the rod (for example periodic strikes in the rod's end with a certain frequency) can lead to torsional vibrations due to the nonlinear parametric interaction of oscillations. In the context of this, the reverse leads to control not only the rotation parameters, but also the direction of rotation, and there are possible torsional oscillations, which can be used in precision mechanics devices.

## 1. Relevance of the problem

Numerous ways of parametric excitation and suppression of oscillations in continuous media [1-14], in particular in liquid ones, are known, for example, the physics of surface phenomena and thin films. Intensive development of modern industry and technology requires the study of parametric excitations (due to the resulting periodic force) or simply parametric oscillations in continuous media: dynamic resistance of elastic systems, fluctuations of plates, shells, rods, oscillations of liquids in vibration vessels, in pipelines, pumps, etc. New trends in modern natural science have appeared in connection with the study of nonlinear processes of various physical, mechanical and chemical nature: thermal hydraulic and magnetohydrodynamic instabilities, disasters, bifurcations, self-oscillation, and so on. Construction of the adequate physical and mathematical models of parametric oscillations of the media and processes occurring in them, the search for ways to effectively influence these processes, and control them - one of the most important directions the of modern continuum mechanics and applied physics.

## 2. Formulation of the problem

This article is devoted to mathematical modeling and analysis of the behavior of parametric nonlinearly coupled parametric longitudinal and torsional oscillations of a rotating rod. Similar tasks are important for many practical applications, in particular, in precision mechanics, printers, and the like. So, at the Institute of Cybernetics of the Academy of Sciences of the USSR (Kyiv) with the participation of the head of the department G.O. Gurvich and leading specialists V.S. Lenchuk, V.V. Bazilevich, P.G. Shishkin a complex of works were executed in the field of development of the elements and systems with the pulsed droplet-jet registration of information; The small-size single-nozzle and multi-nozzle printheads, the microprocessor control systems of high-performance, the monochrome and color printing devices designed for use in small computers, the terminal stations, and the facsimile digital devices were successfully developed and manufactured [15]. In the early 1990s, we bought their cheap printer that was used for high-quality color printing with the conventional cheap ink instead of an expensive printer refueling.

**Mathematical model of nonlinear parametrically coupled longitudinal-torsional rod oscillations**

For the development of thenathematical model for parametric nonlinearly connected longitudinal-torsional oscillations of a rotating rod, the following model of the elastic rod oscillations is considered (figure and the end of paper). The rotating shafts are used as a model for calculating torsional vibrations of the engines. In the theory of the finite displacements, the nonlinear relations that bind them with elastic deformations [16-18] have the following form in the Cartesian coordinate system $x, y, z$:

$$\varepsilon_x = \frac{\partial u}{\partial x} + \frac{1}{2}\left(\frac{\partial u}{\partial x}\right)^2 + \frac{1}{2}\left(\frac{\partial v}{\partial x}\right)^2 + \frac{1}{2}\left(\frac{\partial w}{\partial x}\right)^2,$$

$$\varepsilon_y = \frac{\partial v}{\partial y} + \frac{1}{2}\left(\frac{\partial u}{\partial y}\right)^2 + \frac{1}{2}\left(\frac{\partial v}{\partial y}\right)^2 + \frac{1}{2}\left(\frac{\partial w}{\partial y}\right)^2,$$

$$\varepsilon_z = \frac{\partial w}{\partial z} + \frac{1}{2}\left(\frac{\partial u}{\partial z}\right)^2 + \frac{1}{2}\left(\frac{\partial v}{\partial z}\right)^2 + \frac{1}{2}\left(\frac{\partial w}{\partial z}\right)^2, \quad (1)$$

$$\gamma_{xy} = \frac{\partial u}{\partial y} + \frac{\partial v}{\partial x} + \frac{\partial u}{\partial x}\cdot\frac{\partial u}{\partial y} + \frac{\partial v}{\partial x}\cdot\frac{\partial v}{\partial y} + \frac{\partial w}{\partial x}\cdot\frac{\partial w}{\partial y},$$

$$\gamma_{yz} = \frac{\partial v}{\partial z} + \frac{\partial v}{\partial y} + \frac{\partial u}{\partial y}\cdot\frac{\partial u}{\partial z} + \frac{\partial v}{\partial y}\cdot\frac{\partial v}{\partial z} + \frac{\partial w}{\partial y}\cdot\frac{\partial w}{\partial z},$$

$$\gamma_{xz} = \frac{\partial w}{\partial x} + \frac{\partial u}{\partial z} + \frac{\partial u}{\partial z}\cdot\frac{\partial u}{\partial x} + \frac{\partial v}{\partial z}\cdot\frac{\partial v}{\partial x} + \frac{\partial v}{\partial z}\cdot\frac{\partial w}{\partial x}.$$

In the partial differential equation array (1), the following parameters are: $\varepsilon_x, \varepsilon_y, \varepsilon_z$ – the elements of the normal, and $\gamma_{xy}, \gamma_{xz}, \gamma_{yz}$ – the tangential components of the deformation tensor. Then $\{u, v, w\}$ – the components of the components of the vector of displacements by the corresponding coordinates. In the case of rotation of a circular cylinder of the length $\ell$ with a constant distributed stress along the axis $x$ of the cylinder, while maintaining the shape of the cross-sections:

$$v = -y(1-\cos\varphi) - z\sin\varphi, \qquad w = -z(1-\cos\varphi) + y\sin\varphi. \quad (2)$$

where $\varphi$ is the angle of twisting. Substituting these expressions into the equation of deformation-displacement and putting $\rho = \sqrt{y^2 + z^2}$, we get the following:

$$\varepsilon_x = \frac{\partial u}{\partial x} + \frac{1}{2}\left(\frac{\partial u}{\partial x}\right)^2 + \frac{1}{2}\rho^2\cdot\left(\frac{\partial u}{\partial x}\right)^2, \varepsilon_y = \varepsilon_z = 0, \quad (3)$$

$$\gamma_{xy} = -z\frac{\partial\varphi}{\partial x}, \quad \gamma_{yz} = 0, \quad \gamma_{zx} = y\frac{\partial\varphi}{\partial x}, \quad \gamma = \sqrt{\gamma_{xy}^2 + \gamma_{zx}^2}.$$

A linear relationship between the stresses and deformations is adopted. The longitudinal stress is: $\sigma_x = E\cdot\varepsilon_x\left(1+O(\varepsilon_x^2)\right)$, and the relusting shear stress is:

$$\tau = G\cdot\gamma\cdot\left[1 + \frac{2}{3}\gamma_2\gamma^2 + \frac{4}{9}\gamma_4\gamma^4 + O(\gamma^6)\right], \quad (4)$$

where E is the elastic modulus, $G$ is the displacement module, $\gamma_2, \gamma_4$ are the physical constants of the material that are determined experimentally, $O(\xi)$ is the order magnitude $\xi$, to estimate the errors.

## 3. Control of the parametric oscillations of the rod

In the control of movement of the system (1) - (4) includes elastic restorative forces created by these stresses. Let $F$ be the cross-sectional area, and $r$ is the radius of the cylinder; then the elastic force is:

$$R_1 = \int_F \sigma_x dF = E \cdot F \cdot \frac{du}{dx} + \frac{1}{2} E \cdot F \left(\frac{du}{dx}\right)^2 + \frac{1}{4} r^2 \cdot E \cdot F \left(\frac{d\varphi}{dx}\right)^2 + O\left[\left|\frac{du}{dx}\right|^3 + \left(\frac{d\varphi}{dx}\right)^6\right]. \quad (5)$$

If the polar moment of inertia is introduced $I_0 = 0,5\pi \cdot r^4$ for cross sectional area and the new constants for the material are: $\Gamma_2 = 4/9\gamma^2$ and $\Gamma_4 = 4/9\gamma^4$, then the elastic twisting moment is:

$$R_2 = \int_F \tau \rho dF + R_z = GI_0 \left[\frac{d\varphi}{dx} + r^2 \Gamma_2 \left(\frac{d\varphi}{dx}\right)^3 + r^4 \Gamma_4 \left(\frac{d\varphi}{dx}\right)^5 + O\left[\left(\frac{d\varphi}{dx}\right)^7\right]\right] + R_z, \quad (6)$$

where $R_z$ is an additional twisting moment due to rotate of the forces $\sigma_x$, dF, which are created by interaction of the individual fibers. In this case, the rotation of two planes of the cross-section there are components $\delta_x \rho dF \cdot d\phi/dx$ perpendicular to the radius $\rho$ that create the moment:

$$R_z = \int_F \sigma_x dF \frac{d\varphi}{dx} \rho^2 = EI_0 \frac{d\varphi}{dx} \left\{\frac{du}{dx} + \frac{1}{2}\left(\frac{du}{dx}\right)^2 + \frac{1}{3}r^2\left(\frac{d\varphi}{dx}\right)^2 + O\left[\left|\frac{du}{dx}\right|^3 + \left(\frac{du}{dx}\right)^6\right]\right\}. \quad (7)$$

Longitudinal displacement and the angle of twisting $\varphi$ of the shaft at the free end of the rod (and hence of the disk connected to a rod) satisfy the equation:

$$du/dx = u/\ell, \qquad d\varphi/dx = \varphi/\ell. \quad (8)$$

The mass and moment of inertia of the shaft are neglected in comparison with the mass $\mu$ and moment of inertia $\theta$ of the disk, so that the kinetic energy equals: $T = 0,5\mu(u^1)^2 + 0,5\theta(\phi')^2$, where the stroke marked a differentiation by time.

Next, the Lagrange equation of the second kind for undamplified free oscillations of the rod is written as:

$$\frac{d}{dt}\left(\frac{\partial T}{\partial u'_\upsilon}\right) - \frac{\partial T}{\partial u_\upsilon} + R_0 = 0, \qquad \upsilon = 1,2, \qquad u_1 = u, \qquad u_2 = u, \quad (9)$$

where from with account of (5) - (9) follows:

$$\mu u'' + E \cdot Fu/\ell + 0,5E \cdot Fu^2/\ell^2 + 0,25r^2 E \cdot F\varphi^2/\ell^2 + O(|u|^3 + \gamma^6) = 0, \quad (10)$$

$$\theta\ddot\varphi + GI_0\varphi/\ell + r^2 I_0/\ell^3 = (G\Gamma_2 + 1/3 \cdot E)\varphi^3 + EI_0 u\varphi/\ell^2 + r^4 GI_0 \Gamma_4 \varphi^5/\ell^5 + 0,5EI_0 u^2\varphi/\ell^3 + O(|u^3\varphi| + |\varphi^7|).$$

### The Eigen and parametrically excited oscillations of the system

The second equation in the system (10) describes the Eigen torsional oscillations. Since the longitudinal displacement $u$ is included only together with the angle of twisting $\varphi$, it is convenient for obtaining of the estimations. Neglecting in the first equation of the mathematical model (10) the magni-tudes of order $u^2, \varphi^4$, and in the second - of the order $u^2\varphi, \varphi^5$ of magnitude, respectively, we obtain the equations of motion, which serve as the basis for further research:

$$\mu u'' + E \cdot Fu/\ell + 0,25 r^2 E \cdot Fu^2/\ell^2 \cdot \varphi^2 = 0, \qquad \theta\varphi'' + G\dot{I}_0\varphi/\ell + r^2 \dot{I}_0/\ell^3 (G\Gamma_2 + 1/3 \cdot E)\varphi^3 + E\dot{I}_0 u\varphi/\ell^2 = 0. \qquad (11)$$

Taking into account the first equation of system (11) and dissipative losses in the rod (the coefficient of losses $\gamma$ is equal to the ratio of energy $\Delta W$ absorbed by the elementary volume of the rod during the period of oscillations $T = 2\pi/\omega$ to the potential energy of the elastic deformation W, $\gamma = \Delta W / 2\pi\omega$ ), the following equation is obtained:

$$\theta_s \cdot \varphi'' + G \cdot \dot{I}_0 \varphi/\ell + \left\{ r^2 \dot{I}_0/\ell^3 (G \cdot \Gamma_2 + 1/3 \cdot E) + \frac{(-1) r^2 E^2 \dot{I}_0 \cdot F}{4\ell^4 \sqrt{(E \cdot F/\ell - \mu_s \omega^2)^2 + (E \cdot F\gamma/\ell)^2}} \right\} \varphi^3 = 0. \qquad (12)$$

Here $\theta_s = \theta_d + \theta_r$ – the total moment of Inertia for the rod and disk; $\mu_s$ - the total mass (of the rod and disk); $\mu_s = \mu_d + \mu_r$.

## 4. The nonlinearity of the equations and the resonant regimes

The equations of motion (11), (12) are nonlinear with respect to displacements. The terms with characteristics of the material $\Gamma_2, \Gamma_4$ have a nonlinear relationship between the stresses and deformations, that is, reflect the physical nonlinearity, while the other nonlinear expressions are due to the geometry of deformation. They establish a connection between the angle of rotation and longitudinal displacement. The effect of the nonlinearity of the two types can be compared by the equation (12). For example, for Siemens Martin steel we have $\gamma_2 = -8,5 \cdot 10^4$ and $\Gamma_2 = -3,8 \cdot 10^4$, and therefore in the region far from resonance $\mu_s \omega^2 \neq E \cdot F/\ell$, the factor in front of $\varphi^3$ is mainly determined by a physical nonlinearity. And under the influence of geometric nonlinearity it varies by about 0.002%. But when approaching the resonance ( $\omega^2 \to E \cdot F\ell/\mu_s$ ), the contribution from geometric nonlinearity may increase depending on the value $\gamma$ that can be determined from the area of the loop of the hysteresis of the rod material during its cyclic deformation. The magnitude of the elastic torque can be determined from (6), (7) after the solution of the equation (12), which has an exact solution through the Jacobi elliptic functions.

**The case of long waves and account of the distribution of a system and correlation conditions**

For the long-wave oscillations, if a length of the rod $\ell \sim \lambda$, wave processes from the generator to the disk should be taken into account, that is, the distribution of the system should be taken into account: a shaft (rod) with a disk that interact with the generator, which will change the resonant conditions of the interaction of the generator with the mechanical distributed system. The wave processes in the rod change the load applied to the generator of oscillations.

The terms of coordination will also change. All this can be taken into account within the framework of the linear theory of the wave propagation along the rod. Then instead of $\left[(E \cdot F/\ell - \mu_s \omega^2)^2 + (E \cdot F\gamma/\ell)^2\right]^{-0,5}$ the following yields in the equation (12):

$$\left\{ \left[(E \cdot F/\ell - \mu_s \omega^2) + \mu_s \omega^2 a_1\right]^2 + (E \cdot F\gamma/\ell + \mu_s \omega^2 d_1)^2 \right\}^{-0,5}, \qquad (13)$$

where $a_1, d_1$ - the wave coefficients accounting the properties of the material.

If the rod (wave conductor, wave spreading medium) is subjected to the harmonic oscillation at the generator frequency $\omega$ on one boundary ($x = 0$), and on the other - the free surface (the plane of the disk), then for $a_1, d_1$ results in the following

$$a_1 = -\frac{\alpha \cdot sh(2\alpha l) + \beta \cdot \sin(2\beta \ell)}{\ell(\alpha^2 + \beta^2)[ch(2\alpha\ell) + \cos(2\beta\ell)]}, \quad d_1 = \frac{\alpha \cdot \sin(2\alpha\ell) - \beta \cdot ch(2\alpha\ell)}{\ell(\alpha^2 + \beta^2)[ch(2\alpha\ell) + \cos(2\beta\ell)]}. \quad (14)$$

Here are $\alpha = \frac{\omega}{c_e}\sqrt{\frac{\sqrt{1+\gamma^2}-1}{2(1+\gamma^2)}}$, $\beta = \frac{\omega}{c_e}\sqrt{\frac{\sqrt{1+\gamma^2}+1}{2(1+\gamma^2)}}$. The coefficient $\gamma$ is characterizing an energy dissipation during a cycle of permanent oscillations with the amplitude corresponding to a value of the system's potential energy of elastic body. It can be connected with the logarithmic decrement of oscillation as $\delta$: $\gamma = 2\delta$.

If the concentrator of oscillations is present in a distributed system, then correlation (13) is multiplied by $M^3$, which is the amplifying coefficient of oscillations ($u \sim \varphi^2$). Also, by transition through a resonance $\omega_0^2 = E \cdot F / (\mu\ell)$ (for longitudinal waves), there is a characteristic feature in a change of phase $\psi$ difference between the oscillation (longitudinal) and the exciting force by variation of the frequency of exciting force. This phase difference is always negative so that the oscillations are delaying in a relation to the external exciting force. For example, by $\omega < \sqrt{E \cdot F / (\mu\ell)}$ it is $\psi \to O$, and by $\omega > \sqrt{E \cdot F / (\mu\ell)}$ it is $\psi \to -\pi$. Variation of the $\psi$ from 0 to $-\pi$ is going in a narrow range by frequencies ($\sim \gamma$), close to $\omega_0 = \sqrt{E \cdot F / (\mu\ell)}$. By $\omega = \omega_0$ the phase difference is $\psi = -\pi/2$. In an absence of a friction, the phase change for the excited oscillations by $\pi$ is done abruptly by $\omega = \omega_0$. But with an account of a friction this jump is smoothed.

For the steel, this coefficient with account of physical nonlinearity of a material is $\gamma_2 = 8,5 \cdot 10^4$. And $\Gamma_2 = -3,8 \cdot 10^4$. In a vicinity of a resonance, the equation for $\varphi$ contains the nonlinear term $\sim \varphi^3$ and some corrections as concern to shift in the phases $\psi, r$. The amplifying coefficient depends on consentrator. The corresponding equation is as follows:

$$\theta_s \cdot \varphi''' + GI_0\varphi/\ell + \left\{r^2\dot{i}_0/\ell^3(G \cdot \Gamma_2 + 1/3E) + \frac{(-r^2)E^2\dot{i}_0 FM^3 \cos\varphi}{4\ell^4\sqrt{[(E \cdot F\gamma/\ell - \mu\omega^2) + \mu\omega^2 a_1]^2 + [E \cdot F\gamma/\ell + \mu\omega^2 d_1]^2}}\right\}, \quad (15)$$

where

$$\sin\psi = \frac{E \cdot F\gamma/\ell + \mu\omega^2 d_1}{\sqrt[4]{[(E \cdot F/\ell - \mu\omega^2) + \mu\omega^2 a_1]^2 + [E \cdot F\gamma/\ell + \mu\omega^2 d_1]^2}}. \quad (16)$$

**The possibility for the reverse of the rod's rotations**

For the provement of possibility for the reverse of rotations of the rod and investigation of the conditions for this, the resonance conditions according the expressions (15), (16) are considered. They reveal available change of the sign in front of the $\varphi$, which is of paramount importance being responsible for available reverse of the rod's rotations. This *new phenomenon* found from the built by us mathematical model. It is caused by nonlinear behavior of the parametrically connected longitudinal-torsional oscillations of the rotating rod.

***The special regime exists when parametric action in a form of longitudinal external forced excitation of the oscillations (for example periodic beat at the end of a rod with a given frequency) can result in rotation oscillation due to nonlinear parametric connection of the oscillations.***

The above special regime can be used in a number of different devices, e.g. the ones of the precise mechanics, where the rod must rotate both clockwise and inversely, or suppression of the oscillations in a rod using the beat action in its end, and so on. Thus, it is an interesting new phenomenon with a potential for practical applications in the engineering devices.

In a vicinity of the resonance, the coefficient in front of $\varphi^3$ may be presented in a form:

$$H = \frac{r^2 \dot{I}_0}{\ell^3}\left(G \cdot \Gamma_2 + \frac{1}{3}E\right) + \frac{\left(-r^2 E^2 \dot{I}_0 \cdot F\right) M^3 \cos\psi}{4\ell^4 \sqrt{\left(\mu\omega^2 a_1\right)^2 + \left(E \cdot F\gamma/\ell + \mu\omega^2 d_1\right)^2}} \; . \quad (17)$$

If neglecting the effects of distributed elastic dissipative and inertial properties and inertial wavy part of the system rod-shaft, then

$$H = \frac{r^2 I_0}{\ell^3}\left(G \cdot \Gamma_2 + \frac{1}{3}E - \frac{E \cdot M^3 \cos\psi}{4\gamma}\right). \quad (18)$$

For the resonance, $\cos\psi \to sign\left[\omega - \sqrt{E \cdot F/(\mu\ell)}\right]$ must be, therefore from (18) follows:

$$H = \frac{r^2 \dot{I}_0}{\ell^3}\left(G\Gamma_2 + \frac{1}{3}E + \frac{E \cdot M^3}{4\gamma} sign\left[\omega - \sqrt{\frac{E \cdot F}{\mu\ell}}\right]\right). \quad (19)$$

As far as $\Gamma_2 < 0$, a possibility for a change of the sign of parameter $H$ follows from equation (19) near the longitudinal resonance of the rod. It was assumed in the calculations that $M = 5$. For the coefficient, the data of Academician G.S. Pysarenko with co-authors were used, which are presented in the table:

| Steel grade | E, N/m$^2$ | N/m$^2$ | % |
|---|---|---|---|
| SHX 9 | $2,195 \cdot 9,8 \cdot 10^{10}$ | $0,85 \cdot 9,8 \cdot 10^{10}$ | 0,09(I) |
|  |  |  | 0,04 (II)* |
| ZXIZ | $2,2 \cdot 9,8 \cdot 10^{10}$ | $0,87 \cdot 9,8 \cdot 10^{10}$ | 0,66(I) |
|  | $2,2 \cdot 9,8 \cdot 10^{10}$ | $0,87 \cdot 9,8 \cdot 10^{10}$ | 0,07 (II)*** |
|  | $2,14 \cdot 9,8 \cdot 10^{10}$ | $0,85 \cdot 9,8 \cdot 10^{10}$ | 0,069(III) |
| X17H2 | $2,1 \cdot 9,8 \cdot 10^{10}$ | $0,85 \cdot 9,8 \cdot 10^{10}$ | 0,075(IV)*** |

\* - the steel after annealing ($H_B$=184); - after heating to 840 $^o$C, exposure for 1 hour, hardening in oil, release from 150 $^o$C, keeping for 3 hours, on air ($H_B$ =449) .

\*\* I – given state ($H_B$ = 185);
 II - after thermal processing to the hardness $H_{RC}$ = 45-50;
 III - after thermal processing to the hardness $H_{RC}$ = 50-55;
\*\*\* IV - after thermal processing to the hardness $H_{RC}$ = 40-45;

For example, for the steel grade SHX9 /II/ by $\gamma = 0,0008 = 8 \cdot 10^{-4}$ follows:

$$H = \frac{r^2 I_0}{\ell^3}\left\{0,85 \cdot 9,8 \cdot 10^{10}\left(-8,5 \cdot 10^4\right) + \frac{2,195 \cdot 9,8 \cdot 10^{10}}{3} - \frac{2,195 \cdot 9,8 \cdot 10^{10} \cdot 5^3}{4,8 \cdot 10^{-4}}\right\}, \qquad \omega < \sqrt{\frac{EF}{\mu\ell}}.$$

The main income to $H$ is coming from the terms I and III in the main brackets. By $\omega < \omega_0, H < 0$. Obviously, by $\omega > \omega_0$ the last term in the brackets for $H$ is changing its sign to the opposite one and $H > 0$ becomes by $\omega > \omega_0$. This is a possibility for the reverse of the torsional oscillations of a rod.

**5. The conclusions from the results obnained.**

The developed mathematical model was applied to a study of the nonlinear processes in a system of the interconnected longitudinal-torsional parametric oscillations of the rotating rod and the special regime of a resonance of nonlinear oscillations. The calculation was done, which revealed the correlation between the parameters responsible for the change of a sign in front of the term determining the possibility of a reverse of rotation. Thus, the new phenomenon of the reverse of torsional oscillations due to coupled longitudinal-torsional para-metric oscillations of the rotating rod was discovered and investigated analytically. The phenomenon may have very exciting unique practical applications. The regime of a reverse allows controlling not only rotation of a rod but a change of the rotational direction too.

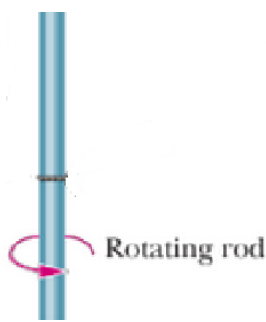

Figure. Rotating rod: to the problem of coupled parametric oscillations